\documentclass{jfm}
\usepackage{graphicx}
\usepackage{epstopdf, epsfig}

\usepackage[colorlinks, citecolor=blue]{hyperref}
\usepackage{xcolor}
\usepackage{soul}
\usepackage{setspace}

\shorttitle{Retracting viscoelastic liquid filaments}
\shortauthor{U. Sen et al.}

\title{The retraction of jetted slender viscoelastic liquid filaments}

\author{Uddalok Sen\aff{1}
  \corresp{\email{u.sen@utwente.nl}},
  Charu Datt\aff{1},
  Tim Segers{\aff{1}$^{,}$\aff{2}},
  Herman Wijshoff{\aff{3}$^{,}$\aff{4}},
  Jacco H. Snoeijer\aff{1},
  Michel Versluis\aff{1},
 \and Detlef Lohse{\aff{1}$^{,}$\aff{5}}
 \corresp{\email{d.lohse@utwente.nl}}}

\affiliation{\aff{1}Physics of Fluids Group, Max Planck Center for Complex Fluid Dynamics, Department of Science and Technology, MESA+ Institute for Nanotechnology, and J. M. Burgers Centre for Fluid Dynamics, University of Twente, 7500 AE Enschede, The Netherlands
\aff{2}BIOS Lab on a Chip Group, Max Planck Center for Complex Fluid Dynamics, Department of Science and Technology, MESA+ Institute for Nanotechnology, and J. M. Burgers Centre for Fluid Dynamics, University of Twente, 7500 AE Enschede, The Netherlands
\aff{3}Department of Mechanical Engineering, Eindhoven University of Technology, 5600 MB Eindhoven, The Netherlands
\aff{4}Canon Production Printing B. V., 5900 MA Venlo, The Netherlands
\aff{5}Max Planck Institute for Dynamics and Self-Organization, 37077, G\"{o}ttingen, Germany}

\begin{document}

\maketitle

\begin{abstract}
Long and slender liquid filaments are produced during inkjet printing, which can subsequently either retract to form a single droplet, or break up to form a primary droplet and one or more satellite droplets. These satellite droplets are undesirable since they degrade the quality and reproducibility of the print, and lead to contamination within the enclosure of the print device. Existing strategies for the suppression of satellite droplet formation include, among others, adding viscoelasticity to the ink. In the present work, we aim to improve the understanding of the role of viscoelasticity in suppressing satellite droplets in inkjet printing. We demonstrate that very dilute viscoelastic aqueous solutions (concentrations $\sim$ 0.003\% wt. polyethylene oxide (PEO), corresponding to nozzle Deborah number $De_{n}$ $\sim$ 3) can suppress satellite droplet formation. Furthermore, we show that, for a given driving condition, upper and lower bounds of polymer concentration exist, within which satellite droplets are suppressed. Satellite droplets are formed at concentrations below the lower bound, while jetting ceases for concentrations above the upper bound (for fixed driving conditions). Moreover, we observe that, with concentrations in between the two bounds, the filaments retract at velocities larger than the corresponding Taylor-Culick velocity for the Newtonian case. We show that this enhanced retraction velocity can be attributed to the elastic tension due to polymer stretching, which builds up during the initial jetting phase. These results shed some light on the complex interplay between inertia, capillarity, and viscoelasticity for retracting liquid filaments, which is important for the stability and quality of inkjet printing of polymer solutions.
\end{abstract}

\begin{keywords}
\end{keywords}

\section{Introduction}

Drop-on-demand inkjet printing is known for its capability of highly-controlled, non-contact deposition of picoliters of liquid material \citep{basaran-2002-aichej, wijshoff-2010-physrep, derby-2010-armaterres, book-hoath, lohse-2022-arfm}. Recent advances have enabled and enhanced the ability to deposit liquids over a wide range of surface tensions and viscosities \citep{castrejonpita-2013-atomizationsprays}. The high degree of reproducibility has led to inkjet printing applications in a diverse array of applications, including text or graphical printing on paper, fabrication of displays in electronics \citep{shimoda-2003-mrsbull}, electronics printing \citep{majee-2016-carbon, majee-2017-carbon}, and in the life sciences \citep{villar-2013-science, daly-2015-intjpharm, simaite-2016-sensactuatorsbchem}. 

A typical inkjet printhead primarily consists of an ink reservoir, a piezo-acoustic transducer, and a dispensing nozzle \citep{wijshoff-2010-physrep}, while the most simple driving waveform is a monopolar trapezoidal pulse \citep{castrejonpita-2008-revsciinstrum}, with a pulse width equal to half the period corresponding to the resonance frequency. Inkjet printheads are usually operated in the \lq pull-push' mode, where the liquid is first pulled into the nozzle during the rise time of the trapezoidal pulse and then pushed out during the fall time of the pulse \citep{fraters-2020-prappl}. This results in the creation of a slender liquid jet of finite length and after pinch-off from the nozzle, a finite liquid ligament with a relatively large head droplet and a long tail. As the ligament is traveling towards the substrate, the tail retracts into the head droplet due to surface tension. However, during such motion, the tail may also break up due to the Rayleigh-Plateau instability \citep{fraters-2020-prappl}. This breakup leads to the formation of satellite droplets, which travel at a velocity lower than that of the head droplet. Thus the head droplet and the satellite droplet(s) reach the substrate at different times, potentially resulting in misalignment and substantially reduced print quality \citep{wijshoff-2010-physrep, derby-2010-armaterres}, and contamination within the print device.

The detrimental effect of the formation of satellite droplets has resulted in an increased focus on the development of techniques to suppress such satellites. These techniques are either based on changing the driving waveform \citep{dong-2006-pof, fraters-2020-prappl} or modifying the properties of the ink. For the latter, since the satellite droplets primarily result from a Rayleigh-Plateau instability of the retracting tail, an obvious choice is to increase the viscosity of the ink in order to stabilize the tail filament \citep{notz-2004-jfm, castrejonpita-2012-prl, driessen-2013-pof, wang-2019-jfm, planchette-2019-prf, anthony-2019-prf}. 

Another strategy \citep{christanti-2002-jrheol} for the suppression of satellite droplet formation is to include polymer additives, which impart viscoelasticity, in the liquid being jetted. Early work has shown that viscoelasticity can stabilize a capillary jet against breakup \citep{goldin-1969-jfm}, and can also suppress satellite droplets when the liquid is jetted by a forced disturbance (such as in inkjet printing) \citep{christanti-2002-jrheol}. \citet{shore-2005-pof} experimentally demonstrated that the addition of polymers to water-based inks suppresses satellite droplet formation in an inkjet printing configuration, while the same observation was also reported in the numerical investigation by \citet{morrison-2010-rheolacta}. \citet{yan-2011-pof} studied the effect of adding polymers, specifically polyethylene oxide (PEO), to water-based inks in inkjet printing, and concluded that the addition of low molecular weight polymers has no significant effect on the overall dynamics of the jetting behavior. \citet{hoath-2012-jrheol} identified experimentally scaling laws relating the maximum jettable concentration to the molecular weight of the polymer additive for polystyrene (PS) in diethyl phthalate (DEP) solutions. For these polymer solutions, it was also reported \citep{hoath-2014-jnnfm} that there is a delay in the pinch-off of the liquid filament from the nozzle, as compared to a purely Newtonian ink. 

Despite the recent surge in the investigations of inkjet printing with polymer solutions, and the extensive literature \citep{bazilevskii-1990-erc, anna-2001-jrheol, amarouchene-2001-prl, clasen-2006-jfm, eggers-2020-jfm, zhou-2018-prf, zhou-2020-pof} on the thinning and breakup of polymeric jets and pendant drops, there seems to be a dearth in the quantitative understanding of the fundamental physical mechanisms responsible for the suppression of satellite droplets in such conditions. Here, we quantitatively study the breakup (or retraction) of the long ligaments produced during (polymeric) ink jetting and the subsequent satellite droplet formation (or suppression). In particular, we identify the operating range where no satellite droplets are observed. In this range, the temporal retraction behavior of the jetted ligament length is characterized, and explained by a simple theoretical model. The model also quantifies the forces responsible for the suppression of satellite droplets, and demonstrates reasonable agreement with the experimental observations. 

The paper is organized as follows: \S~\ref{sec:methods} describes the experimental procedure. In \S~\ref{sec:jetting} the experimental results for different values of the control parameters are shown, culminating in the phase diagram (Fig.~\ref{fig:regime map}). In \S~\ref{sec:ret exp}  we present detailed and quantitative experimental measurements of the filament retraction, which are theoretically explained in \S~\ref{sec:ret theory}. The paper ends with conclusions and an outlook in \S~\ref{sec:conclusions}.

\section{Experimental procedure} \label{sec:methods}

Along the lines of previous work \citep{christanti-2002-jrheol, shore-2005-pof, yan-2011-pof}, PEO (average molecular weight $\simeq$ 10$^{6}$ a.u., Sigma-Aldrich, henceforth referred to as PEO1M) was chosen to be the polymer additive in this work. Aqueous solutions of PEO1M, of concentrations ($c$, by mass) ranging from 0.001\% to 0.009\%, were prepared by adding the required amount of polymer powder to purified water (Milli-Q). For each polymer concentration, a 100 mL solution is made at first and all experiments were carried out using the same solution. The required polymer powder amount was measured using a precision laboratory balance (Secura 224-1S, Sartorius) with an accuracy of 0.1~mg, and then added to purified water (Milli-Q). Each solution was stirred with a magnetic stirrer for 24 h prior to use in order to ensure homogeneity of concentration. This results in a high degree of repeatability of the experiments (as evident later from the small error bars in Fig.~\ref{fig:contraction length}a). In the present experiments, we wanted to have polymer solutions that can act as Boger fluids~\citep{james-2009-arfm}, i.e., solutions whose shear viscosities are independent of the shear rate. This allows for the development of the simplified model, shown in \S~\ref{sec:ret theory}, where the shear rate dependence on viscosity is not taken into account. Dilute aqueous solutions of PEO1M act as Boger fluids \citep{cooperwhite-2002-jnnfm}. Thus PEO1M was chosen as the polymer to be tested. This was also verified independently in the present work. The rheological characterization of the test liquids was performed on a rotational rheometer (MCR 502, Anton-Paar) with a cone-and-plate configuration (1$^{\circ}$ angle, 50 mm diameter, and mean gap of 0.1 mm). The measured viscosities are mentioned in Table~\ref{tab:prop}, where $c^{\ast}$ is the critical overlap concentration~\citep{clasen-2006-jrheol}. That being said, the results of the present work, in particular the simplified theoretical model (\S~\ref{sec:ret theory}), can be extended to any Boger fluid composed of a different polymer. This can be easily facilitated by characterizing the rheological properties of the test fluid such as relaxation time ($\lambda$), solvent ($\eta_{s}$) and polymer ($\eta_{p}$) viscosities, along with knowing the density ($\rho$) and surface tension ($\gamma$). These material properties can then be incorporated into the theoretical approach (\S~\ref{sec:ret theory}). The relaxation times were measured from the extensional thinning of the liquid filaments in a pendant droplet configuration \citep{deblais-2018-prl, deblais-2020-jfm} (see also relaxation time measurement in a CaBER (Capillary Breakup Extensional Rheometer) device \citep{bazilevskii-1990-erc, anna-2001-jrheol, amarouchene-2001-prl, clasen-2006-jfm}). The surface tension was measured by the pendant droplet method in an optical contact angle measurement and contour analysis instrument (OCA15, Dataphysics). It was noted that the surface tensions and the viscosities for the solutions of different polymer concentrations remained practically unchanged, while an appreciable change was observed in the relaxation time measurements (also observed by \citet{yan-2011-pof}). It is known from literature~\citep{noskov-2000-jphyschemb} that changing the PEO concentration can change the surface properties of the solutions. However, in the very narrow concentration range (0.001\% -- 0.009\% by wt.) used in the present work, the surface properties are practically independent of the polymer concentration~\citep{noskov-2000-jphyschemb}. Hence the surface properties of the solution were not taken into account. Nonetheless, it would be an interesting study to see how the surface properties of the solution affects the filament stability, and some recent numerical studies~\citep{wee-2020-prl, kamat-2020-jfm, wee-2021-jfm} have already provided an indication of significant effects of surface properties. However, a detailed experimental study investigating the effects of the surface properties is beyond the scope of the present work. A recent study~\citep{walls-2015-pre} also suggests that the viscosity of the outer gas can play a role in the breakup of a liquid filament. However, the focus of the present work is to study the effect of viscoelasticity on the retraction of a liquid filament. In that context, the effect of the outer gas viscosity is also not within the scope of the current study. Therefore, experiments with different surrounding gases were not performed.

\begin{figure}
\centering
\includegraphics[width=\textwidth]{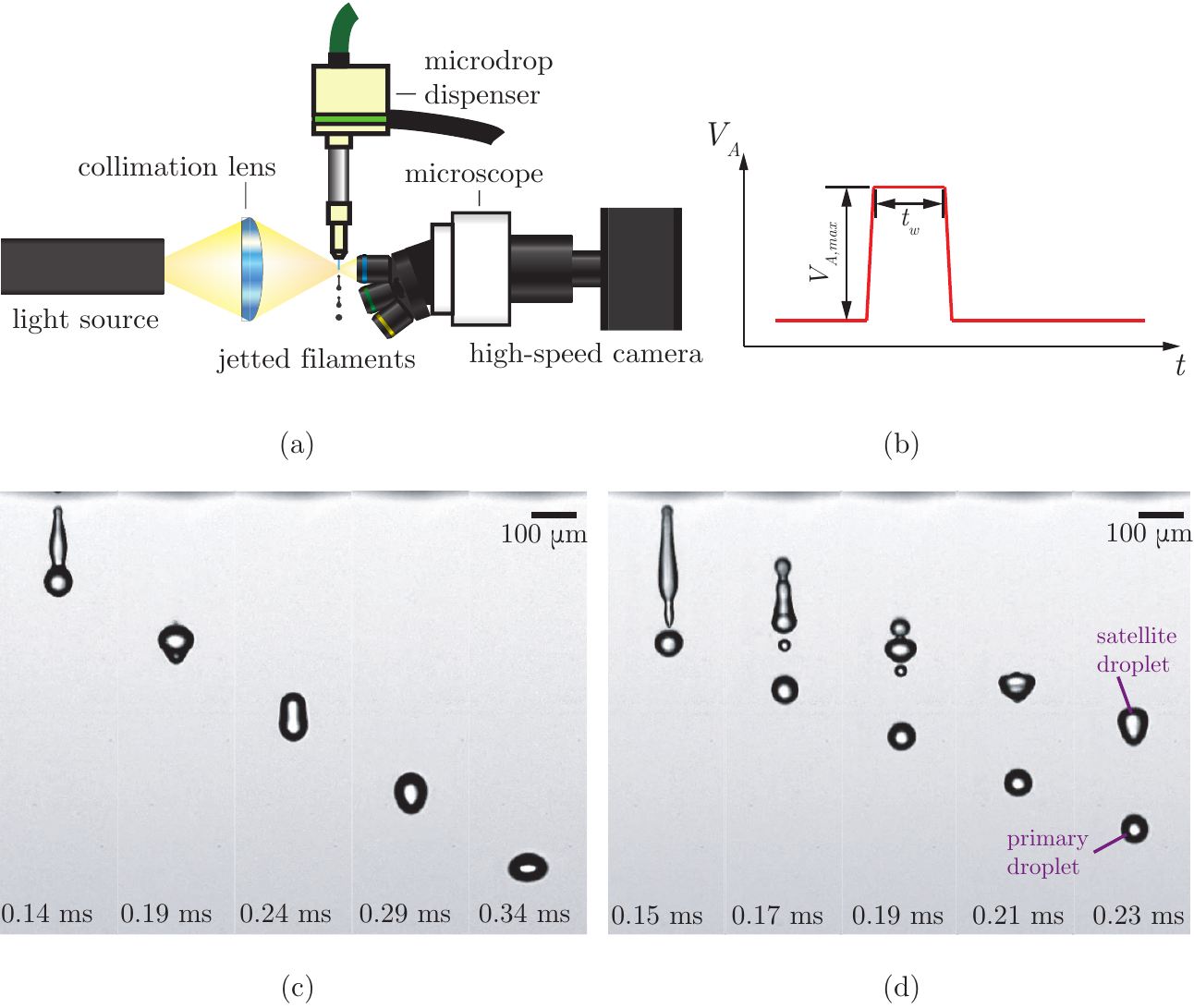}
\caption{(a) Schematic of the experimental setup. (b) Typical trapezoidal pulse used for actuation. (c) Jetting with water at $V_{A, max}$ = 50 V, $t_{w}$ = 40 $\mu$s results in a single droplet. (d)~Jetting with water at  $V_{A, max}$ = 60 V, $t_{w}$ = 40 $\mu$s results in the formation of a primary droplet and a satellite droplet.}
\label{fig:setup}
\end{figure}

\begin{table}
  \begin{center}
\def~{\hphantom{0}}
\caption{Salient properties of the PEO1M solutions used in the present work.}
  \begin{tabular}{ccccccc}
      \hline
      $c$ [wt\%] & $c/c^{\ast}$ & $\eta_{s}$ [mPa.s] & $\eta$ [mPa.s] & $\eta_{p}$ [mPa.s] & $\lambda$ [$\mu$s] & $\gamma$ [N/m]\\[3pt]
      \hline
       0.001 & 0.007 & 0.89 & 0.91 & 0.02 & $\simeq$ 0 & 0.07\\
       0.003 & 0.022 & 0.89 & 0.91 & 0.02 & 28.2 & 0.07\\
       0.005 & 0.037 & 0.89 & 0.91 & 0.02 & 44.72 & 0.07\\
       0.007 & 0.052 & 0.89 & 0.91 & 0.02 & 77.98 & 0.07\\
       0.009 & 0.067 & 0.89 & 0.91 & 0.02 & 99.38 & 0.07\\
       \hline
  \end{tabular}
   \label{tab:prop}
  \end{center}
\end{table}

The schematic of the experimental setup is shown in Fig.~\ref{fig:setup}a (a photograph of the setup has been provided as part of the supplementary information). A microdrop dispenser (AD-K-501, Microdrop Technologies GmbH), with a nozzle inner diameter of 50 $\mu$m was used to generate the liquid ligaments. The dispenser consists of a cylindrical piezoacoustic transducer glued around a glass capillary connected to a fluid reservoir. A detailed description of such a dispenser can be found in \citet{dijskman-1984-jfm, fraters-2021-arxiv}. The piezoacoustic element is driven by an electrical pulse supplied from an arbitrary waveform generator (WW1072, Tabor Electronics) and amplified (50$\times$) by a high-voltage amplifier (WMA-300, Falco Systems). A typical driving electrical pulse is trapezoidal in shape \citep{wijshoff-2010-physrep}, as shown in Fig.~\ref{fig:setup}b, where $V_{A}$ denotes voltage and $t$ denotes time. The rise and fall times of the pulse are kept constant in the present experiments at 1~$\mu$s, while the amplitude ($V_{A,max}$) and the pulse width ($t_{w}$) were varied in the ranges 50~--~75 V and 30 -- 50 $\mu$s, respectively. The lower limits of these ranges are set by the minimum driving required for jetting, while the upper limits are dictated by a bubble entrainment phenomenon associated with meniscus destabilization at strong driving conditions \citep{fraters-2019-prappl, fraters-2021-arxiv}. The shape of the supply waveform was verified by an oscilloscope (DPO 4034B, Tektronix). The liquid in the dispenser was supplied from a transparent plastic syringe (5 mL, Becton-Dickinson) connected via a flexible plastic PEEK tubing (Upchurch Scientific). The microdrop dispenser was driven continuously at a drop-on-demand frequency of 100 Hz to minimize selective evaporation from the nozzle, thereby ensuring a constant liquid composition. High-speed imaging of the jetting behavior was performed at 10$^{5}$ fps (frames-per-second), with a 600 ns exposure time, by a high-speed camera (Fastcam SA-Z, Photron) connected to a modular microscope (U-ECA and BXFM-F, Olympus) with multiple objectives (5$\times$, 10$\times$, and 20$\times$, MPlanFL N, Olympus). This allowed for a spatial resolution of as low as 1 $\mu$m per pixel. The experiments were illuminated by a LED light source (70\% intensity, KL 2500 LED, Schott), with the light beam being collimated onto the imaging plane by a collimation lens (Thorlabs). The light source used in the present work is a cold light source, and was only used at 70\% of its maximum intensity. Furthermore, the light source would only be switched on during image capturing, and remained switched off when the camera was not capturing any frames (such as during changing or refilling the liquid, and saving the captured frames on the computer hard drive). Thus it is expected that the temperature of the liquid in the device does not change appreciably due to the presence of the light source. The dispenser and the camera were triggered simultaneously (with nanosecond precision) by a programmable pulse-delay generator (BNC 575, Berkley Nucleonics Corp.). The captured images were further analyzed using an OpenCV-based Python script developed in-house. The script utilized a Canny edge detection method to find the edges of the jetted ligament outside the nozzle and that of the retracting meniscus within the nozzle. Knowing the pixel location of the edges, one can find out the top-most and bottom-most point of the jetted ligament, and thus its length (defined as $L(t)$ in Fig.~\ref{fig:contraction length}a). Similarly, knowing the top-most point of the retracting meniscus within the nozzle, $z_{m}$ (Fig.~\ref{fig:meniscus}d), is tracked.

\section{Jetting liquids with polymer additives} \label{sec:jetting}

The jetting behavior with water at $V_{A, max}$= 50 V, $t_{w}$ = 40 $\mu$s is shown in Fig.~\ref{fig:setup}c (and Movie 1 of  the supplementary information). Here, and in all subsequent figures, the time $t$ = 0 is defined as the moment when the dispenser is triggered. At that moment, the recording with the camera is also started. A detailed description of the inkjet droplet formation can be found in \citet{book-hoath, wijshoff-2010-physrep}. Briefly, a liquid ligament is formed following pinch-off from the nozzle. As the ligament propagates, the tail retracts due to capillarity, and merges with the head droplet to form a single droplet. When the pulse amplitude ($V_{A, max}$) is further increased to 60 V, a longer ligament is initially produced (as seen in Fig.~\ref{fig:setup}d and Movie 1 of the supplementary information); however, in this case, the tail does not retract into the head droplet. Instead, the ligament breaks up into multiple smaller droplets through the Rayleigh-Plateau instability. These smaller droplets in Fig.~\ref{fig:setup}d then merge to form a larger satellite droplet, which never coalesces with the primary droplet owing to the satellite's lower speed.

\begin{figure}
\centering
\includegraphics[width=\textwidth]{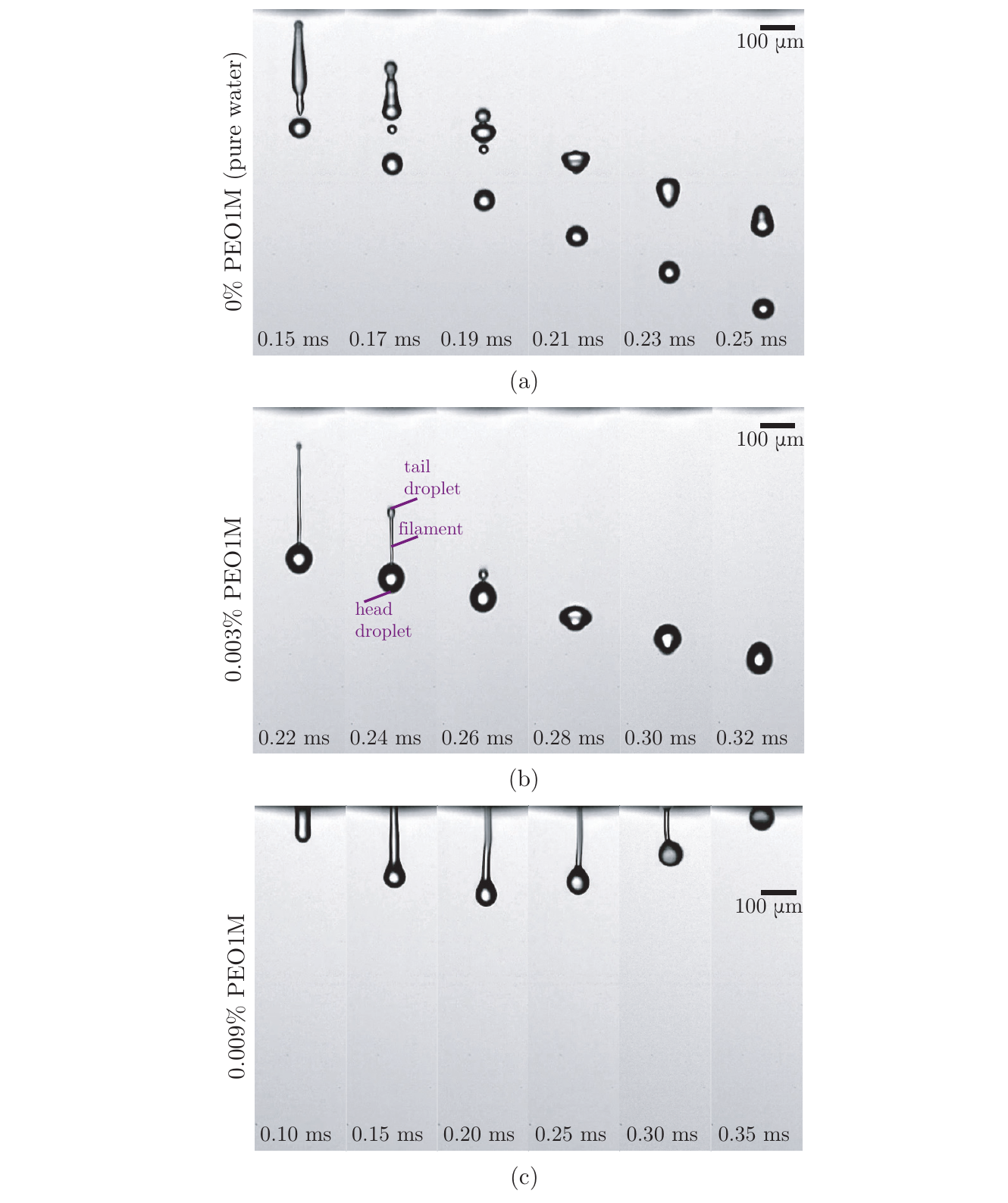}
\caption{(a) Jetting with water at $V_{A, max}$ = 60 V, $t_{w}$ = 40 $\mu$s results in the formation of a satellite droplet (`satellite' regime). (b) Jetting with 0.003\% PEO1M solution ($\lambda$~=~28.2~ms, $c/c^{\ast}$~=~0.022, $De_{n}$~=~1.91) at $V_{A, max}$ = 60~V, $t_{w}$ = 40 $\mu$s results in the suppression of satellite droplet formation (`no satellite' regime). (c) A 0.009\% PEO1M solution ($\lambda$~=~99.38~ms, $c/c^{\ast}$~=~0.067, $De_{n}$~=~6.75) at $V_{A, max}$ = 60 V, $t_{w}$ = 40 $\mu$s results in no detachment of droplets from the nozzle (`no jetting' regime).}
\label{fig:jetting}
\end{figure}

Next we add the polymer (PEO1M) to the fluid being jetted. The typical jetting behavior for three different concentrations is shown in Fig.~\ref{fig:jetting}. In all the three cases shown in the figure, the driving conditions are kept constant at $V_{A, max}$ = 60 V, $t_{w}$~=~40~$\mu$s; only the concentration of PEO1M is varied. The jetting behavior with water at this condition is shown in Fig.~\ref{fig:jetting}a, reproduced from Fig.~\ref{fig:setup}d. For the 0.003\% PEO1M solution (Fig.~\ref{fig:jetting}b and Movie 2 of the supplementary information), the jetting behavior shows a stark difference as compared to that of water. Now the jetted ligament consists again of a spherical head droplet and a long slender tail, similar to the pure water case. Note that the tail here is significantly thinner than the one for the weaker driving case with water (Fig.~\ref{fig:setup}c). A small spherical tail droplet is also observed in this case, which grows in size as the tail retracts towards the head droplet. When the head and the tail droplet are sufficiently close, they merge to form a single droplet, without any satellite droplets being formed. Thus, the addition of a very small quantity of PEO1M (0.003\% by mass) is sufficient to suppress the formation of satellite droplets. Addition of the long chain polymer imparts viscoelasticity to the aqueous solution. The slender tail acts like a stretched filament, being forced by elasticity (and capillarity) to retract its length. This results in the retraction of the whole liquid filament without any intermediate break up. By observing the time stamps, it can also be identified that the pinch-off from the nozzle happens at a later time for the 0.003\% PEO1M solution (0.22 ms) as compared to that for water (0.15 ms). This has been reported in \citet{hoath-2014-jnnfm} also, albeit for a different polymer-solvent combination. We note that the suppression of satellite droplet formation by the addition of viscoelasticity comes at the cost of some jetting velocity. For example, the velocity of the jetted 0.003\% PEO1M droplet (Fig.~\ref{fig:jetting}b) was measured to be 3.11~m/s, as compared to the 3.33 m/s measured for the primary droplet in the pure water case (Fig.~\ref{fig:jetting}a). A hypothesis for this decrease in velocity is as follows: In a piezo-acoustic fluid dispenser such as the one used in the present work, the electrical pulse input to the transducer induces a mechanical deformation of the piezo-actuator, resulting in a pressure pulse in the liquid within the dispenser. A detailed description of the associated mechanisms can be found in \citet{dijskman-1984-jfm, wijshoff-2010-physrep}. From an energetics point of view, the pressure pulse imparts an additional energy in the fluid volume. For a Newtonian liquid, this energy is converted into the surface energy of the jetted droplet and its kinetic energy. However, for a viscoelastic liquid, there is an additional elastic energy associated with stretching the liquid ligament. Since the droplet sizes for the Newtonian and the viscoelastic cases are almost the same, resulting in similar surface energies (as surface tension remains unchanged  at such low polymer concentration, as seen in Table~\ref{tab:prop}), the resulting kinetic energy for the viscoelastic droplet is lower. This results in a lower jetting velocity. However, this is only a qualitative understanding of the physical phenomenon, and a detailed quantitative understanding is beyond the scope of the present work.

\begin{figure}
\centering
\includegraphics[width=\textwidth]{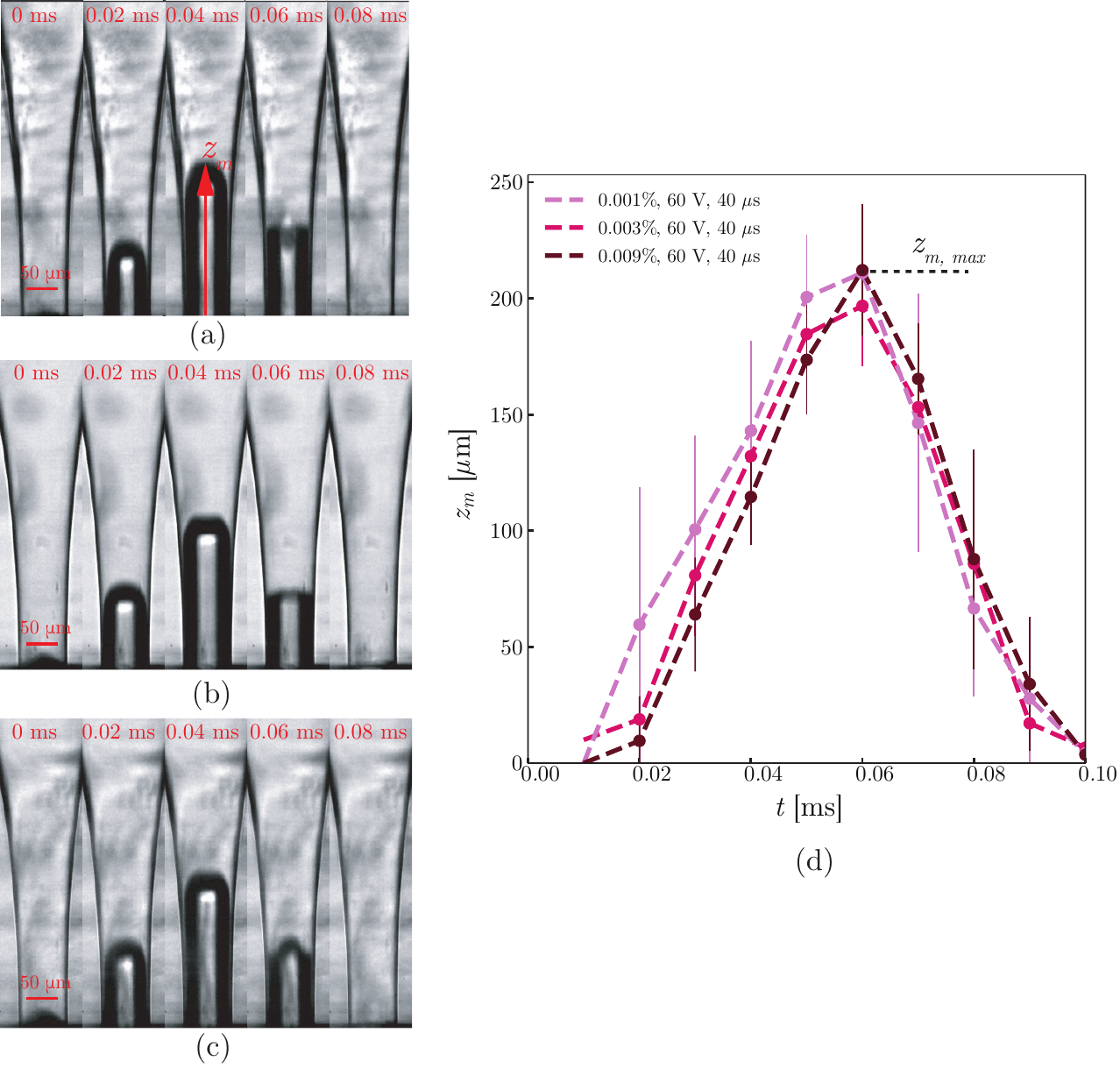}
\caption{Motion of the liquid meniscus inside the nozzle, at $V_{A, max}$ = 60 V, $t_{w}$ = 40 $\mu$s, for (a) 0.001\% PEO1M, (b) 0.003\% PEO1M, and (c) 0.009\% PEO1M solutions, basically showing no difference. (d) The variation of the meniscus position ($z_{m}$) with time ($t$). For the same driving condition (both $V_{A, max}$, $t_{w}$ kept constant), the maximum meniscus position ($z_{m, max}$) is independent of the polymer concentration. The dashed lines are guides for the eyes.}
\label{fig:meniscus}
\end{figure}

If the polymer concentration is further increased to 0.009\% by mass (still with the same driving), the jetting behavior again changes dramatically, as seen in Fig.~\ref{fig:jetting}c (and Movie 2 of the supplementary information). In this case, a filament of the liquid, while still connected to the liquid inside the nozzle, appears downstream of the nozzle exit. However, now the additional polymer content has increased the elasticity to such an extent that the filament does not pinch-off from the nozzle. Instead, it retracts back into the nozzle, thus suppressing jetting altogether. \citet{morrison-2010-rheolacta} refer to such observations as the \lq bungee jumper' in their numerical study. Obviously, for inkjet printing or similar droplet deposition applications, this phenomenon is undesirable, as no droplet is being produced. 

\begin{figure}
\centering
\includegraphics[width=0.85\textwidth]{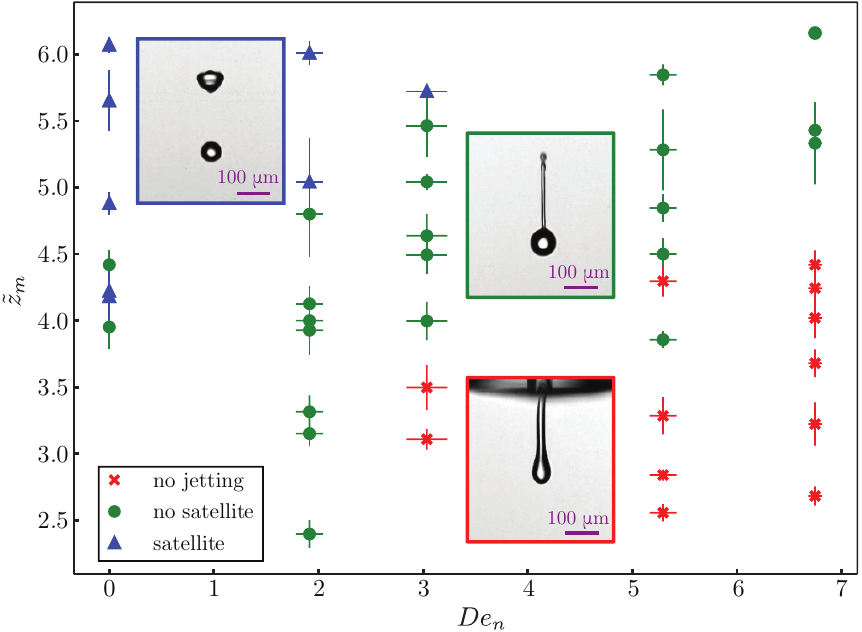}
\caption{Regime map based on the nozzle Deborah number ($De_{n}$) and the non-dimensional maximum meniscus position ($\tilde{z}_{m}$). The insets show the typical jetting (or lack of) behavior observed in each regime, namely satellite formation (blue markers) for low $De_{n}$ and high $\tilde{z}_{m}$, no satellite formation (green markers) for intermediate $De_{n}$ and $\tilde{z}_{m}$, and no jetting (red markers) for high $De_{n}$ and low $\tilde{z}_{m}$. Each datapoint represents approximately 25 experiments.}
\label{fig:regime map}
\end{figure}

The vastly different jetting behaviors observed in Fig.~\ref{fig:jetting} suggest that, for a given driving condition, there exists a concentration range where jetting without the formation of satellites is observed, and on either side of that range the jetting deteriorates, namely by the occurrence of satellites for lower concentrations, or by suppressing the jetting altogether at high concentrations. Hence it is imperative to delineate these three regimes in an appropriate two-dimensional parameter space, with suitable variables representing the solution concentrations and the driving conditions. 

The choice of the control parameter to represent the driving condition is not straightforward, as both pulse amplitude ($V_{A, max}$) and pulse width ($t_{w}$) are varied in the present experiments. Since the optically transparent microdrop dispenser is operated in the `pull-push' mode, the motion of the liquid meniscus inside the nozzle, just prior to filament pinch-off, can be observed in order to gauge the effect of the driving condition. The meniscus motion for three different solution concentrations, at the same driving condition, are shown in Fig.~\ref{fig:meniscus} (and Movie 3 of the supplementary information). A detailed description of such motion, and how it is affected by changing the driving conditions, can be found in \citet{fraters-2021-arxiv}. The position of the meniscus, $z_{m}$, is tracked as shown in Fig.~\ref{fig:meniscus}a, and its temporal evolution is plotted in Fig.~\ref{fig:meniscus}d. From Fig.~\ref{fig:meniscus}d, it is observed that the maximum meniscus position, $z_{m, max}$, is the same for the three different solution concentrations. In other words, $z_{m, max}$, is only affected by the pulse amplitude and pulse width \citep{fraters-2021-arxiv}, and not by the amount of polymer present in the solutions used in the present work. Thus, $z_{m, max}$ can be considered to be a suitable variable to represent the effect of the driving conditions. It is non-dimensionalized with the nozzle diameter, $d_{n}$, as
\begin{equation}
\tilde{z}_{m} = \frac{z_{m,max}}{d_{n}} .
\end{equation}

The second dimensionless parameter for the two-dimensional parameter space must be related to the composition of the jetted liquid. The solution concentration is associated to its relaxation time, $\lambda$. With increasing polymer concentration, $\lambda$ increases \citep{deblais-2018-prl, deblais-2020-jfm}. The relaxation time is non-dimensionalized with the capillary time, resulting in the nozzle Deborah number, $De_{n}$, defined as
\begin{equation}
De_{n} = \frac{\lambda}{t_{\gamma, n}} ,
\end{equation}
with
\begin{equation}
t_{\gamma, n} = \left( \frac{\rho d_{n}^{3}}{8 \gamma} \right)^{1/2} ,
\end{equation}
where $\rho$ is the density of the fluid and $\gamma$ is the surface tension. 

The regime map in the $\tilde{z}_{m}$-$De_{n}$ parameter space is shown in Fig.~\ref{fig:regime map}. As expected from the observations made in Fig. \ref{fig:jetting}, the desirable `no satellite' (green markers) regime lies in the middle of the phase space, flanked by the `satellite' (blue markers) and `no jetting' (red markers) regimes. For a particular polymer concentration (constant $De_{n}$), a stronger driving (higher $\tilde{z}_{m}$) is required to overcome the elastic effects and traverse from the \lq no jetting' regime to the \lq no satellite' regime. However, if the driving is too strong, the stabilizing effect of elasticity on the retracting tail is lost, resulting in the tail to break, forming satellites. 

\section{Filament retraction: experiments} \label{sec:ret exp}

\begin{figure}
\centering
\includegraphics[width=\textwidth]{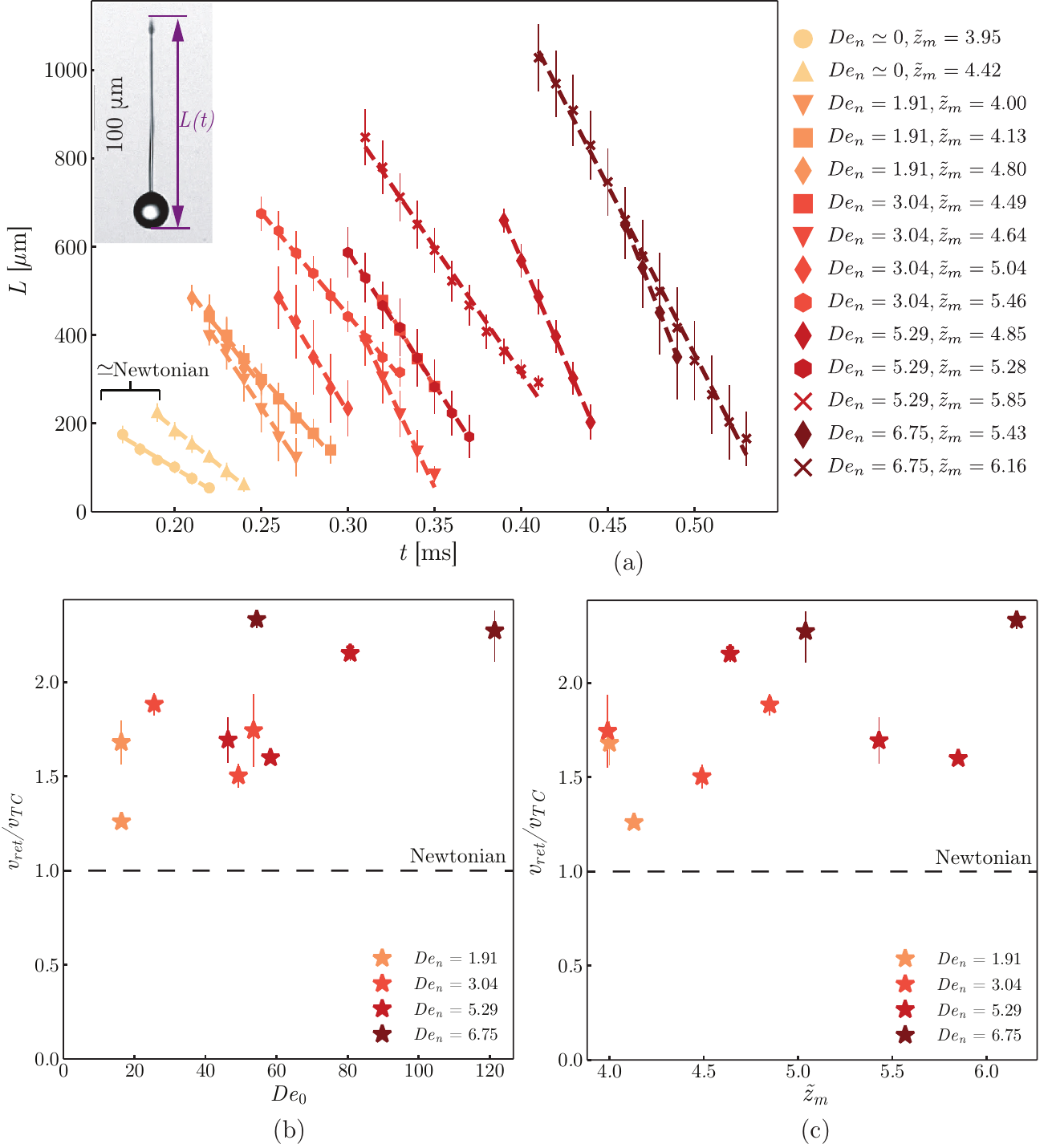}
\caption{(a) Temporal evolution of the length $L(t)$ of the traveling liquid filaments with time $t$. The inset shows a typical measurement. The dashed lines are linear fits. (b) Variation of the ratio of the filament retraction velocity ($v_{ret}$, measured from the experiments) to the corresponding Newtonian Taylor-Culick velocity ($v_{TC}$), plotted against the filament Deborah number, $De_{0}$. The dashed line indicates the Newtonian behavior, $v_{ret} = v_{TC}$. (c) Variation of $v_{ret}/v_{TC}$ plotted against the maximum meniscus position ($\tilde{z}_{m}$). The dashed line indicates Newtonian behavior, $v_{ret} = v_{TC}$.}
\label{fig:contraction length}
\end{figure}

The instantaneous retraction length, $L(t)$, is measured between the extremities of the head droplet and the tail droplet, as shown in the inset of Fig. \ref{fig:contraction length}. The temporal evolution of $L(t)$ is plotted in Fig. \ref{fig:contraction length}a for different solution concentrations ($De_{n}$) and driving conditions ($\tilde{z}_{m}$). The retraction length appears to decrease linearly with time even for the viscoelastic liquids ($De_{n} \neq 0$) (dashed lines in Fig. \ref{fig:contraction length}a); a trend that is expected for Newtonian liquids ($De_{n} \simeq 0$) \citep{planchette-2019-prf}. The retraction velocity that is extracted from the slopes of these curves is approximately constant in time for each dataset shown in Fig.~\ref{fig:contraction length}a, but different from that of a Newtonian filament (see Fig.~\ref{fig:contraction length}a), where it is equal to the Taylor-Culick velocity \citep{keller-1983-pof, hoepffner-2013-jfm, pierson-2020-prf}: 
\begin{equation}
v_{TC} = \left( \frac{\gamma}{\rho R_{0}} \right)^{1/2} ,
\end{equation}
where $R_{0}$ is the radius of the retracting filament. In the present experiments, $R_{0}$ is measured at a location on the contracting filament close to the head droplet, and at a time instant $t = t_{0}$ after pinch-off from the nozzle. In certain experiments, post pinch-off, spurious effects such as out-of-plane oscillations were observed. These were adjudged to be initial transients, and removed from the datasets plotted in Fig.~\ref{fig:contraction length}a. $t_{0}$ thus corresponds to the first datapoint of each dataset in Fig.~\ref{fig:contraction length}a. The error bars in Fig.~\ref{fig:contraction length} represent $\pm$~one standard deviation across approximately 25 experiments for each dataset. The ratio of the retraction velocities ($v_{ret}$) of the viscoelastic filaments in the current experiments (measured from the linear fits in Fig.~\ref{fig:contraction length}a) to the corresponding Newtonian Taylor-Culick velocity ($v_{TC}$) is plotted in Fig.~\ref{fig:contraction length}b against the filament Deborah number, defined as
\begin{equation}
De_{0} = \frac{\lambda}{t_{\gamma}} ,
\label{eq:De0}
\end{equation}
where
\begin{equation}
t_{\gamma} = \left( \frac{\rho R_{0}^{3}}{\gamma} \right)^{1/2} .
\end{equation}
In the present experiments, $De_{0} /De_{n} \sim \mathcal{O}(10)$. Furthermore, $R_{0}$ was observed to vary over a very narrow range (4 -- 13 $\mu$m) in the present experiments, resulting in more than one $De_{0}$ value in some cases for a given $De_{n}$. Figure~\ref{fig:contraction length}b shows that the viscoelastic filaments have a higher retraction velocity than the Newtonian ones. Moreover, it can be observed from Fig.~\ref{fig:contraction length}b that for the same $De_{0}$, there are more than one $v_{ret}$ value. This suggests that the relaxation time, $\lambda$, is not the only parameter affecting the retraction of these filaments; the retraction is also affected by the driving conditions at which the solution is jetted. To characterize the effect of the driving conditions on $v_{ret}$, we plot $v_{ret}/v_{TC}$ as a function of $\tilde{z}_{m}$ for different $De_{n}$ in \mbox{Fig.~\ref{fig:contraction length}c}. However, no clear trend is observed in the plot. This is probably due to the complex flow profile associated with the motion of the meniscus, and is essentially a limitation of the experimental setup.

\section{Filament retraction: theoretical model} \label{sec:ret theory}

\begin{figure}
\centering
\includegraphics[width=\textwidth]{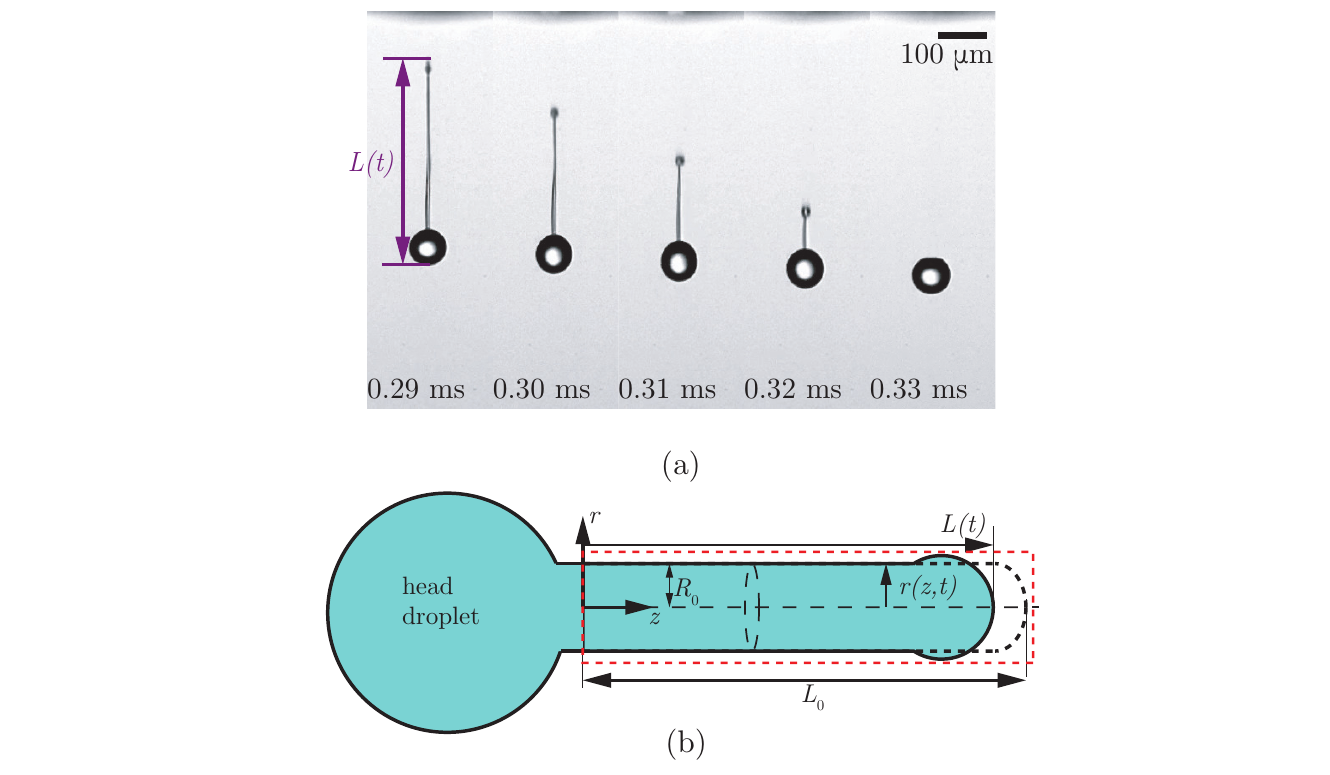}
\caption{(a) The evolution of a jetted 0.005\% PEO1M ligament produced at $V_{A, max}$ = 60 V, $t_{w}$ = 40 $\mu$s. (b) Schematic of the geometry for the theoretical model, also clarifying the employed notation.}
\label{fig:theoretical model}
\end{figure}

In order to identify the role of viscoelasticity on the retraction velocity of liquid filaments, a simplified theoretical model is proposed. The retraction dynamics of viscoelastic liquid films have been studied in other geometries \citep{evers-1997-prl, kdv-1999-pre, villone-2017-jnnfm, tammaro-2018-langmuir, villone-2019-jnnfm}, but not for a slender liquid filament. We follow the lines of \citet{pierson-2020-prf} for Newtonian liquid filaments, but now account for the viscoelasticity due to the polymers. A careful examination of the retraction phenomenon (Fig.~\ref{fig:theoretical model}a) reveals that during the retraction, the size of the head droplet does not change noticeably ($<$ 3\%), while the slender tail is pulled towards the head droplet. During this retraction, the spherical tail droplet grows in size as the tail length decreases. This behavior is modeled by the geometry shown in Fig.~\ref{fig:theoretical model}b. 

At time $t$ = $t_{0}$, the tail in Fig.~\ref{fig:theoretical model}a is modeled as a long cylindrical filament with a rounded end (shown by the black dashed lines in Fig. \ref{fig:theoretical model}b), having radius $R_{0}$ and length $L_{0}$. We consider an axisymmetric coordinate system that is co-moving with the head droplet (the head droplet moves with a constant velocity in the present experiments), which implies that the fluid within the tail is initially at rest in this coordinate system. One can then proceed via a momentum balance over the control volume that is indicated by the red dashed rectangle. The surface and elastic tensions pull the filament towards the head droplet. For $t > t_{0}$, the rounded edge of the filament therefore starts to retract in the negative $z$-direction, with an instantaneous length $L(t)$. As the tail retracts, liquid from the tail feeds the tail droplet \citep{savva-2009-jfm, pierson-2020-prf}, resulting in an increase of its size. Below we will use that most of the liquid momentum is localized inside the tail droplet, as it moves inwards with an instantaneous velocity $dL/dt$. Note that $L(t)$ is defined slightly differently to the definition used in the experimental data. However, since the diameter of the head droplet does not change appreciably, the following analysis holds well for describing the retraction. 

We formalize these ideas using the slender jet approximation ($R_{0} \ll L$), within that the mass and momentum conservations can be written as \citep{eggers-1993-prl, shi-1994-nature, eggers-1994-jfm, clasen-2006-jfm}
\begin{equation}
\frac{\partial r^2}{\partial t} + \left(r^2 v\right)' = 0,
\label{eq:continuity}
\end{equation}
and 

\begin{equation}
\rho \left( \frac{\partial v}{\partial t}  + v v' \right) = -\gamma \kappa'
+ \frac{1}{r^{2}}  \left( 3 \eta_{s} r^{2} v' \right)' 
+ \frac{1}{r^{2}} \left( r^{2} (\sigma_{zz} - \sigma_{rr}) \right)' .
\label{eq:momentum}
\end{equation}
Here $v(z,t)$ and $r(z,t)$, respectively are the axial velocity and the filament radius, prime denotes a derivative along $z$, while $\eta_{s}$ is the solvent viscosity, and $\kappa$ is the curvature of the filament, given by
\begin{equation}
\kappa =  \frac{1}{r(1 + r'^2)^{1/2}}  - \frac{r''}{(1 +r'^2)^{3/2}} .
\label{eq:curvature}
\end{equation}
The viscoelasticity is accounted for by $\sigma_{zz}$ and $\sigma_{rr}$, the components of the polymer stress tensor $\boldsymbol{\sigma}$, for which a separate constitutive equation needs to be specified \citep{book-bird}. In the slender jet geometry, the predominant stretching and viscoelastic contribution is along the $z$-direction, so $\sigma_{rr}$ can be omitted in the remainder \citep{clasen-2006-jfm, book-eggers}. 

In order to perform the control volume analysis, one can bring the slender jet equations to a conservative form by multiplying Eq.~(\ref{eq:continuity}) by $\rho v$ and Eq.~(\ref{eq:momentum}) by $r^{2}$, and then adding them up to obtain \citep{clasen-2006-jfm, eggers-2008-repprogphys}: 

\begin{equation}
 \frac{\partial (\rho r^{2} v)}{\partial t} +  \left(\rho r^{2}v^{2}\right)' =
 \left[ r^2 \left( \gamma K + 3 \eta_{s}  v' + \sigma_{zz}\right) \right]' ,
\label{eq:conservative}
\end{equation}
where 

\begin{equation}
K =  \frac{1}{r(1 + r'^2)^{1/2}}  + \frac{r''}{(1 +r'^2)^{3/2}} .
\label{eq:curvature}
\end{equation}
The right hand side of Eq. (\ref{eq:conservative}) can readily be integrated from $z=0$ to $z=L(t)$, over the control volume in Fig. \ref{fig:theoretical model}b. We integrate the equation using the same assumptions as in \citet{pierson-2020-prf}: (i) $r$ vanishes at $z = L(t)$, as that is the tip of the filament; (ii)~the filament radius is uniform ($r=R_0$ and $r'=0$) at $z$ = 0, which is at an arbitrary location close to the head droplet; (iii) the fluid is at rest ($v=v'=0$)  at $z = 0$. Defining the total momentum $P = \pi \rho \int_{0}^{L(t)} r^{2} v dz$, we can then indeed integrate Eq.~(\ref{eq:conservative}) from $z = 0$ to $z = L(t)$ as \citet{savva-2009-jfm, pierson-2020-prf}:

\begin{equation}
\frac{dP}{dt} =  - \pi \left( \gamma R_{0} + R_{0}^{2} \sigma_{zz} \vert_{z=0}  \right) .
\label{eq:momentum1}
\end{equation}
We recover the anticipated momentum balance, with capillary and elastic forces pulling the liquid tail towards the head drop.

Now, the retraction process is feeding the tail droplet \citep{savva-2009-jfm, pierson-2020-prf}, while the fluid between the head and the tail droplets remains at rest \citep{pierson-2020-prf}. Therefore, $P$ is essentially the momentum of the tail droplet $P = M_T dL/dt$ (assuming that the fluid velocity inside the droplet is constant \citep{pierson-2020-prf}). The mass $M_T(t)$ of the tail drop increases over time by the mass flow rate $- \pi \rho R_{0}^{2} dL/dt$, such that 

\begin{equation}
M_{T}(t)  = \pi \rho R_{0}^{2} (L_{0} - L(t)) + 2 \pi \rho R_{0}^{3}/3.
\end{equation}
In the analysis that follows, we omit the initial mass ($2 \pi \rho R_{0}^{3}/3$) of the edge of the filament \citep{pierson-2020-prf}, which is negligible in the experimental data with which we compare our theoretical calculations. Integration of Eq. (\ref{eq:momentum1}) in time then gives 

\begin{equation}
P =
\pi \rho R_0^2 (L_0-L)\frac{dL}{dT} =  - \pi \left( \gamma R_{0}T +  R_{0}^{2} \int_0^T \sigma_{zz}(\bar t) \vert_{z=0} d\bar t \right),
\label{eq:momentumbis}
\end{equation}
where we have introduced a change of variable with $T = t- t_{0}$. For a Newtonian fluid ($\sigma_{zz}=0$), this equation can be integrated to $(L_0-L)^2 = \frac{\gamma}{\rho R_0}T^2$ and one recovers a retraction with a constant (Taylor-Culick) velocity \citep{keller-1983-pof, hoepffner-2013-jfm, pierson-2020-prf}. It is clear that the presence of elastic stress will offer an extra contribution that speeds up the retraction, as observed in experiments. Importantly, however, the relaxation of $\sigma_{zz}(T)$ will lead to a nonlinear evolution of $L(T)$, so that the retraction velocity is no longer constant. 

To close the problem, we need a constitutive relation for the polymeric stress. Here we use the Oldroyd-B fluid that has been successfully used to describe the thinning of jets \citep{clasen-2006-jfm, eggers-2020-jfm}. In terms of the conformation tensor $\mathbf A$, the  stress is then given by a constitutive relation \citep{book-bird}

\begin{equation}
\boldsymbol{\sigma} = \frac{\eta_p}{\lambda} (\mathbf A  - \mathbf I),
\end{equation}
where $\eta_p$ is the polymer viscosity. In the Oldroyd-B fluid, the conformation tensor evolves by a linear relaxation dynamics, which in the slender jet approximation reads \citep{fontelos-2004-jnnfm}

\begin{equation}
\frac{\partial A_{zz}}{\partial T} + v \frac{\partial A_{zz}}{\partial z} = \frac{1}{\lambda} + \left( 2 \frac{\partial v}{\partial z} - \frac{1}{\lambda} \right) A_{zz} .
\label{eq:constitutive1}
\end{equation}
Using the same assumptions used for deriving Eq.~(\ref{eq:momentum1}) and using that $A_{zz} \gg 1$ (large stretching of polymer chains along the axis), Eq.~(\ref{eq:constitutive1}) can be reduced to

\begin{equation}
\frac{\partial A_{zz} \vert_{z=0} } {\partial T} = - \frac{1}{\lambda}  A_{zz} \vert_{z=0} .
\label{eq:constitutive2}
\end{equation}
The polymer stress then follows as 

\begin{equation}
A_{zz}(T) \vert_{z=0} = A_{0} e^{-T/\lambda}  \quad \Rightarrow \quad \sigma_{zz}(T) \vert_{z=0} = \frac{\eta_p}{\lambda} A_0 e^{-T/\lambda} .
\label{eq:decay}
\end{equation}
The initial condition $A_0=A_{zz}(z=0, T=0)$ is not determined from the present analysis, but is an inherited condition from the jetting phase, where the polymers are deformed by the stretching flow.  

\begin{figure}
\centering
\includegraphics[width=\textwidth]{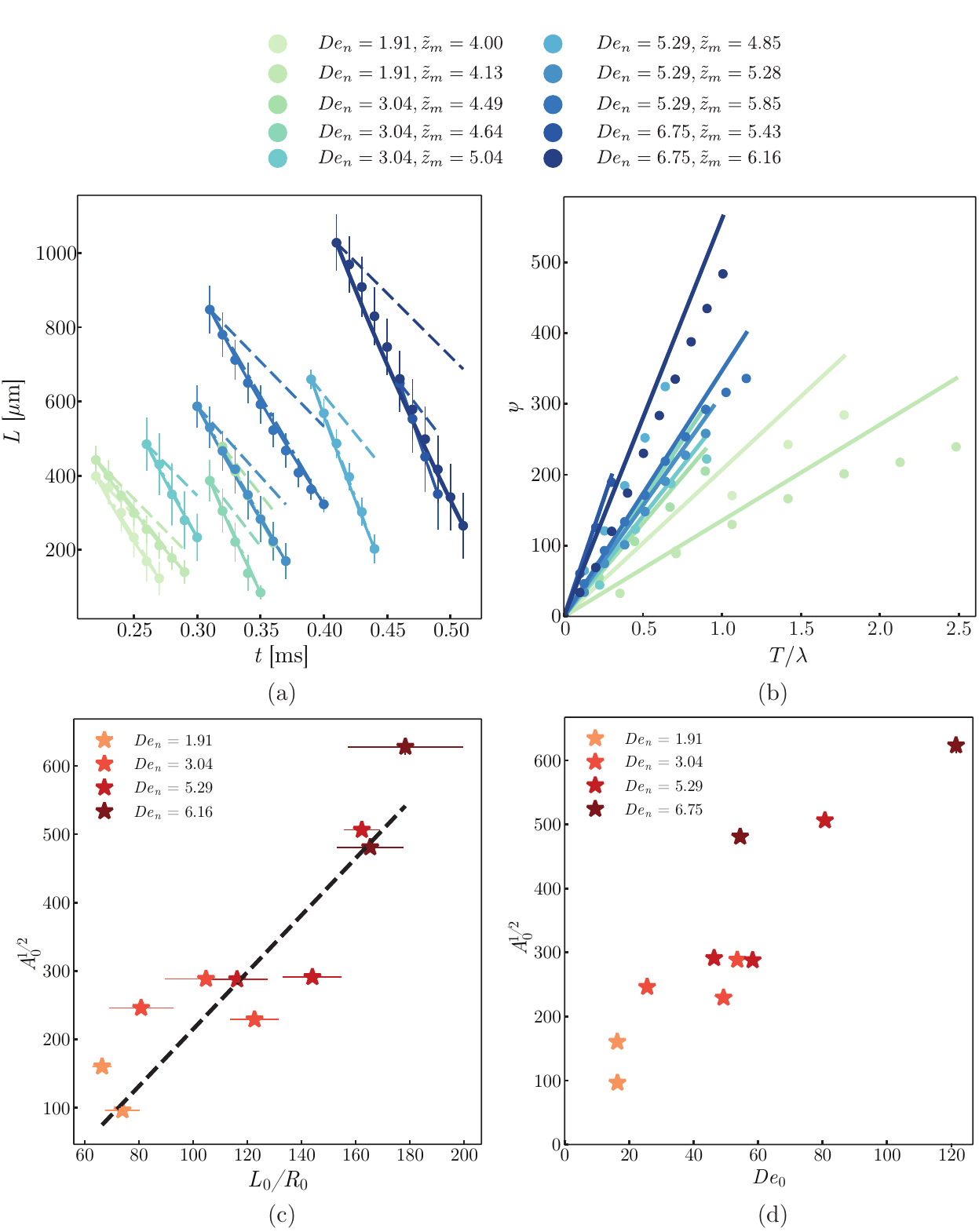}
\caption{Filament length $L(t)$ as function of time: (a) Fitting of the theoretical predictions (solid lines) with experimental observations (discrete datapoints); the dashed lines indicate the corresponding Newtonian (Taylor-Culick) behavior. (b) Linear behavior at small $T/\lambda$ as predicted by Eq.~(\ref{eq:conlength_slope}); the discrete datapoints correspond to the experiments while the solid lines indicate the prediction from Eq.~(\ref{eq:conlength_slope}) valid for small times $T/\lambda < 1$. (c) Variation of the fitted $A_{0}^{1/2}$ with the initial aspect ratio ($L_{0}/R_{0}$). The dashed line representing a linear fit. (d)~Variation of $A_{0}^{1/2}$ with $De_{0}$.}
\label{fig:fitting}
\end{figure}

We now return to Eq. (\ref{eq:momentumbis}) with the polymer stress given by Eq. (\ref{eq:decay}), so that

\begin{equation}
\left( L_{0} - L \right) \frac{dL}{dT} = - \frac{\gamma}{\rho R_{0}} T - \frac{\eta_{p}}{\rho}  A_0 \left( 1 - e^{-T/\lambda} \right) .
\label{eq:velocity1}
\end{equation}
This can be integrated with the initial condition $L(0) = L_{0}$, to yield the variation of the contracting length with time, given by

\begin{equation}
(L_0-L)^2 =  v_{TC}^2 T^{2} + 2 \frac{\eta_{p}}{\rho} A_{0} \lambda \left( e^{-T/\lambda} - 1 + \frac{T}{\lambda} \right) .
\label{eq:conlength}
\end{equation}
This is the central result of the analysis. Although the resulting $L(T)$ is nonlinear, the variation of $L$ with $T$  appears nearly linear (see result plotted in Fig.~\ref{fig:fitting}a). To highlight the effect of viscoelasticity, the corresponding Newtonian Taylor-Culick prediction for each dataset is also shown in Fig.~\ref{fig:fitting}a by dashed lines. It can be clearly observed from Fig.~\ref{fig:fitting}a that viscoelastic retraction (discrete data points from experiments and solid lines from fitting Eq.~(\ref{eq:conlength})) is faster than the corresponding Newtonian Taylor-Culick retraction. Given the nearly linear appearance of $L(T)$, it is therefore instructive to expand Eq.~(\ref{eq:conlength}) for early times $T/\lambda\ll 1$, which gives

\begin{equation}
(L_0-L)^2 =  \left( v_{TC}^2 +  \frac{\eta_{p}A_{0} }{\lambda \rho}   \right) T^{2} . 
\label{eq:conlengthexpand}
\end{equation}
This illustrates the enhanced retraction velocity $\left(v_{TC}^{2} + \frac{\eta_{p} A_{0}}{\lambda \rho} \right)^{1/2}$ during the initial stage. A result of this form can be obtained even more generally, beyond the assumptions underlying the Oldroyd-B fluid. Namely, evaluating the stress integral at short time in Eq. (\ref{eq:momentumbis}), one finds the initial retraction velocity 

\begin{equation}
v_{ret} = \left(v_{TC}^2 + \frac{\sigma_{zz}(z=0, T=0)}{\rho} \right)^{1/2},
\end{equation}
incremented by elastic stress that is initially in the filament. 

Finally, one may use Eq.~(\ref{eq:conlength}) to estimate $A_{0}$ in our experiments, which is otherwise difficult a priori. The fitting was performed by matching $L_{0}$ (at $T$ = 0) and $L$ from the experiments at the final $T$ instant for each data set, as shown in Fig.~\ref{fig:fitting}a by the continuous lines. We rewrite Eq. (\ref{eq:conlengthexpand}) as 

\begin{equation}
\left( \frac{\rho}{\lambda \eta_{p}} \left( \left( L - L_{0}  \right)^{2} - \frac{\gamma}{\rho R_{0}} T^{2}  \right) \right)^{1/2} = A_{0}^{1/2} \frac{T}{\lambda} .
\label{eq:conlength_slope}
\end{equation}
When plotting the left hand side of Eq.~(\ref{eq:conlength_slope}), expressed as $\psi$, against $T/\lambda$, a straight line through (0,0) is expected at small values of $T/\lambda$, with the slope depicting $A_{0}^{1/2}$. This is observed in Fig.~\ref{fig:fitting}b for the experimental data (discrete datapoints), with deviations from linear behavior (continuous lines) observed at larger $T/\lambda$ values. 

Now, $A_{0}=A_{zz} (z=0, T=0)$ is the axial component of the conformation tensor $\mathbf{A}$, which itself is defined as $\mathbf{A} = \langle \vec{X} \vec{X} \rangle/X_{e}^{2}$ \citep{book-bird}, where $\vec{X}$ is the stretched length of each individual polymer molecule, and $X_{e}$ its equilibrium length. Here, the polymer molecules are pictured to be two spherical beads connected by a spring. Therefore, one can argue that $A_{0}^{1/2}$ is proportional to the stretched length of the polymer molecules at $t = t_{0}$ (or $T = 0$) and $z = 0$. The stretched length of the polymer is proportional to the local polymer stretching, $\epsilon_{l}$, while the initial aspect ratio, $L_{0}/R_{0}$, of the filament may be proportional to the stretching of the filament, $\epsilon_{g}$. It may be expected that $\epsilon_{l}$ and $\epsilon_{g}$ are correlated under strong axial tension for a slender liquid filament. Hence, $A_{0}^{1/2}$ can be assumed to be linearly correlated to $L_{0}/R_{0}$. The variation of $A_{0}^{1/2}$ with $L_{0}/R_{0}$ is plotted in Fig.~\ref{fig:fitting}c; the dependence is not inconsistent with the assumed linear behavior. The values of $A_{0}^{1/2}$ obtained from fitting with the experiments are $\mathcal{O}(100)$, and one may wonder whether the finite extensibility of polymers (ignored in the Oldroyd-B model) may play a role. Determining the finite extensibility from rheological experiments is a challenge. \citet{lindner-2003-physicaa} report, for higher molecular weight PEO (2 $\times$ 10$^{6}$ a.u. and 4 $\times$ 10$^{6}$ a.u.), maximum polymer stretched lengths in the range of $\mathcal{O}(10)$ to $\mathcal{O}(100)$. The deviations in $A_{0}^{1/2}$ from the assumed linear trend might therefore be attributed to the limitations of the Oldroyd-B model. We remark, however, that any analysis with nonlinear constitutive relations will come with additional (unknown) fitting parameters. In addition, there are other factors that may play a role in the retraction dynamics such as polydispersity and multiple relaxation time scales of the polymer molecular chains \citep{entov-1997-jnnfm, wagner-2005-prl}, non-uniform radius of the filament along the axial direction, non-axisymmetric effects at the nozzle exit \citep{vdmeulen-2020-prappl}, and wetting effects at the nozzle exit \citep{beulen-2007-expfluids, dejong-2007-apl}, which we have not considered in the present study. Figure~\ref{fig:fitting}c also suggests that $A_{0}^{1/2}$ depends on the polymer concentration, which is expected since the higher the polymer concentration, the longer is the relaxation time of the solution, resulting in the ability of the polymers to be stretched longer. Hence, intuitively, one can come to the realization that $A_{0}^{1/2}$ increases with the polymer concentration, thus the polymer relaxation time $\lambda$, and subsequently $De_{0}$. This is also observed in Fig.~\ref{fig:fitting}d. However, the exact trend for the variation of $A_{0}^{1/2}$ with $De_{0}$ is rather difficult to predict in the present experiments. We note that this may be due to the complex flow profile associated with the motion of the meniscus (Fig.~\ref{fig:contraction length}c), and a simplified approach to estimate the average elongation rate of the polymer macromolecules from the motion of the meniscus itself is unsatisfactory. This can be ascribed to being a limitation of the experimental setup, and more controlled experiments in a simplified geometry can shed light into the elongation rate of the macromolecules (thus $A_{0}$) prior to the commencement of retraction.

We further clarify the applicability of two of the assumptions used for developing the theoretical model: In the present experiments, the filament radius in Fig.~\ref{fig:theoretical model}a is not uniform but decreases along its distance from the head droplet. However, given the optical resolution of our imaging system, it is extremely difficult to characterize this variation. Hence in the theoretical model in \S~\ref{sec:ret theory}, this variation has not been taken into account. Furthermore, as highlighted by \citet{vdbos-2014-prappl}, the fluid velocity inside the droplet produced by an inkjet printhead is not constant. However, the primary goal of the present work was to study the retraction of a viscoelastic filament. A simplified model \S~\ref{sec:ret theory} for that purpose was developed, where it was assumed, for simplicity, that the fluid velocity inside the droplet is constant. Despite these aforementioned simplifications, the model was found to describe the experimentally-observed retraction speeds being faster than the Newtonian Taylor-Culick velocity reasonably well.

The regime map (Fig.~\ref{fig:regime map}) indicates a transition from the no satellite to satellite regime at low $De_{n}$ (0 -- 3) and high $\tilde{z}_{m}$ (4 -- 6). The exact location of this transition probably can be estimated by comparing the retraction timescale obtained from the theoretical model in the present work with the timescale associated with pinch-off. A starting point for such an analysis could be the recent work by \citet{eggers-2020-jfm}. Furthermore, the theoretical model described in the present work is based on the Oldroyd-B model, which cannot capture pinch-off. One would need to consider nonlinear models such as FENE-P to suitably describe pinch-off. Such nonlinear models require additional (unknown) fitting parameters. However, a detailed calculation is beyond the scope of the present work, and has been left as a future exercise.

\section{Conclusions and outlook} \label{sec:conclusions}

The present work demonstrates that adding small amounts of long chain polymers to water-based inks can result in the suppression of satellite droplets formation in inkjet printing. These polymers impart viscoelasticity to the liquid being jetted, resulting in stabilization of the slender finite-length filament against a Rayleigh-Plateau instability. Due to the action of both capillarity and viscoelasticity, the tail droplet is pulled towards the head droplet, while being connected by a thin liquid thread, resulting in the formation of a single droplet without any satellites. However, if the polymer concentration is too high, no droplet is jetted as the increased elasticity prevents pinch-off from the dispensing nozzle. A regime map in terms of the experimental operating parameters was identified, thus delineating the `satellite', `no satellite', and `no jetting' regimes. Moreover, it was observed that, in the `no satellite' regime, the filament essentially retracts linearly with time, with the retraction velocity being higher than the Newtonian Taylor-Culick velocity. A simple theoretical model was developed to derive the retraction velocity for a slender viscoelastic liquid filament, which works reasonably well in modeling the experimental observations. These results are helpful in revealing the complex interplay between inertia, capillarity, and viscoelasticity during the retraction of slender liquid filaments, and are highly useful to predict the behavior during inkjet printing of polymer solutions.

The novelty of the present work does not lie in demonstrating the suppression of satellite droplets by the addition of polymers. Such suppression behavior has been shown before (see \citet{christanti-2002-jrheol, shore-2005-pof, morrison-2010-rheolacta, yan-2011-pof, hoath-2012-jrheol, hoath-2014-jnnfm}). However, a quantitative understanding of the physical mechanism  for the suppression of satellite droplets by the addition of viscoelasticity to the ink has been lacking in the scientific literature. We attempt to address these knowledge gaps through this present manuscript. Systematic experiments have been performed over a range of polymer concentrations and driving conditions that result in observations ranging from satellite droplets to no satellites and ultimately to the suppression of jetting altogether. To the best of our knowledge, a similar investigation has not been hitherto reported. The retraction behavior of viscoelastic filaments reported in this manuscript is also new; in particular the higher retraction speeds as compared to the Newtonian Taylor-Culick velocity. Furthermore, the manuscript presents a simplified theoretical model that explains the apparently linear retraction behavior and shows reasonable agreement with the experimentally-observed retraction speeds being faster than the Newtonian Taylor-Culick velocity. Until now there has not yet been such a comparison between experiment and theory, which we provide, though the complexity of the problem unfortunately does not allow for a quantitative one-to-one comparison. \\

\noindent{\textbf{Acknowledgements}}

U. S. is thankful to Maziyar Jalaal for stimulating discussions, and Maaike Rump for drawing the schematic of the experimental setup. \\
 
\noindent{\textbf{Funding}}
 
The support from an Industrial Partnership Programme of the Netherlands Organisation for Scientific Research (NWO), cofinanced by Canon Production Printing B. V., University of Twente, and Eindhoven University of Technology is acknowledged. C. D. and J. H. S. acknowledge support from NWO VICI Grant No. 680-47-632. \\

\noindent{\textbf{Declaration of interests}}

The authors report no conflict of interest. \\

\noindent{\textbf{Supplementary information}} \label{Movie }

Supplementary information is available at (URL to be inserted by publisher). \\

\noindent{\textbf{Author ORCID}}

U. Sen \href{https://orcid.org/0000-0001-6355-7605}{https://orcid.org/0000-0001-6355-7605};  

C. Datt \href{https://orcid.org/0000-0002-9686-1774}{https://orcid.org/0000-0002-9686-1774};

T. Segers \href{https://orcid.org/0000-0001-5428-1547}{https://orcid.org/0000-0001-5428-1547};  

H. Wijshoff \href{https://orcid.org/0000-0002-2120-0365}{https://orcid.org/0000-0002-2120-0365};

J. H. Snoeijer \href{https://orcid.org/0000-0001-6842-3024}{https://orcid.org/0000-0001-6842-3024};

M. Versluis \href{https://orcid.org/0000-0002-2296-1860}{https://orcid.org/0000-0002-2296-1860};  

D. Lohse \href{https://orcid.org/0000-0003-4138-2255}{https://orcid.org/0000-0003-4138-2255}. \\

\bibliographystyle{jfm}
\bibliography{Bibliography}

\end{document}